# Unwanted electroless zinc plating on current collectors in zinc air batteries

Vinoba Vijayaratnam[1], Harald Natter[1], Samuel Grandthyll[2], Jens Uwe Neurohr[2], Karin Jacobs[2], Frank Müller[2], Rolf Hempelmann[1]

[1] Physical Chemistry, Saarland University, 66123 Saarbrücken, Germany
[2] Experimental Physics, Saarland University, 66123 Saarbrücken, Germany

E-mail: v.vijayaratnam@gmx.de

## 1. Abstract

The occurrence of metallic film deposition without external power supply on the copper current collector of a zinc air battery half-cell containing zinc slurry is investigated. Therefore, test specimens of miscellaneous materials representing the current collector are immersed in a commercial available zinc slurry as well as an in self-prepared zinc slurry. In case of copper and metals which are more noble (silver and gold), a coating on the respective specimen is obtained. An element mapping of the coated copper specimen is performed by means of scanning electron microscopy (SEM) and energy dispersive x-ray spectroscopy (EDX) identifying that the coating layer consists of zinc and oxygen. In order to clarify the crystal structure and the exact composition, focused ion beam (FIB) and x-ray photoelectron spectroscopy (XPS) measurements are applied, proving that the layer consists of elemental zinc, which, is merely oxidized on the surface. Finally, a reaction mechanism for the reversible zinc film deposition is proposed.

## 2. Introduction

Due to the inevitable transition from a fossil fuel based economy to a clean energy economy, a cost-effective technology is needed to solve the problems of the discontinuous production of regenerative energy from e.g. solar, wind and hydroelectric sources. Besides pumped hydroelectric storage, compressed air energy storage or flywheels, electrochemical systems such as batteries are suitable candidates, especially for flexible and scalable energy storage on an industrial scale.[1-11] Also, due to the rising environmental problems, many governments are discussing the possibility to ban new registrations of petrol-powered cars and only allow new registrations of electric vehicles by 2030.[12] Because of this strong global incentive to develop electrical vehicles, the demand for batteries in mass production is increasing in the last years. The batteries should alleviate not only the greenhouse gas emission but also reduce the foreign oil dependence.[13] One of the most promising candidates for these applications is the zinc-air battery due to its high specific energy, low material cost, environmental performance and good safety, compared to conventional batteries, sodium-sulphur batteries, redox-flow batteries, other metal-air batteries or even lithium-ion batteries.[14] The zinc-air battery can be classified between traditional batteries and fuel cells, consisting of a negative zinc electrode and an air-breathing positive oxygen electrode coupled by a suitable electrolyte.[1] Already known to the scientific community since the late 19$^{th}$ century,[15] the zinc-air battery still holds the greatest promise for future energy applications due to its high theoretical energy density of 1086 Wh kg$^{-1}$ including oxygen (about five times higher than lithium-ion batteries) and its very low potential manufacturing cost of < 10 \$ kW$^{-1}$ h$^{-1}$ (about two orders of magnitude lower than lithium-ion batteries).[16,1] Despite its early start and great potential, still some challenging technical problems have yet to be resolved like the formation of potassium carbonate plugging the pores of the gas electrode, the evolution of hydrogen as a side reaction, the inefficiency of available catalysts for the air electrode, the non-uniform zinc dissolution during discharge, the growth of zinc dendrites during recharge and the lack of satisfactory bifunctional air catalysts.[17-22]

The zinc air battery contains three parts: an anode, a cathode and a separator, as shown in Figure 1. Zinc slurry is used as anode side. Zinc slurry consists of small zinc particles with 30 wt% potassium hydroxide solution. The air electrode (GDE) is utilized as cathode side, which is divided into a gas diffusion layer (GDL) and a catalytic active layer. The electrons, generated from the oxidation of zinc slurry, are transported through an external electrical conductor to the cathode, where the electrons are used for the reduction of oxygen from the air to hydroxide ions in the alkaline electrolyte. This process is called as three phase reaction: with oxygen as the gas phase, electrolyte as the liquid phase and the catalyst as the solid phase.[18,17] Through the separator, the hydroxide ions migrate to the anode, where they react with zinc cations to hydroxo-complexes and zinc oxide, where they complete the cell reaction.

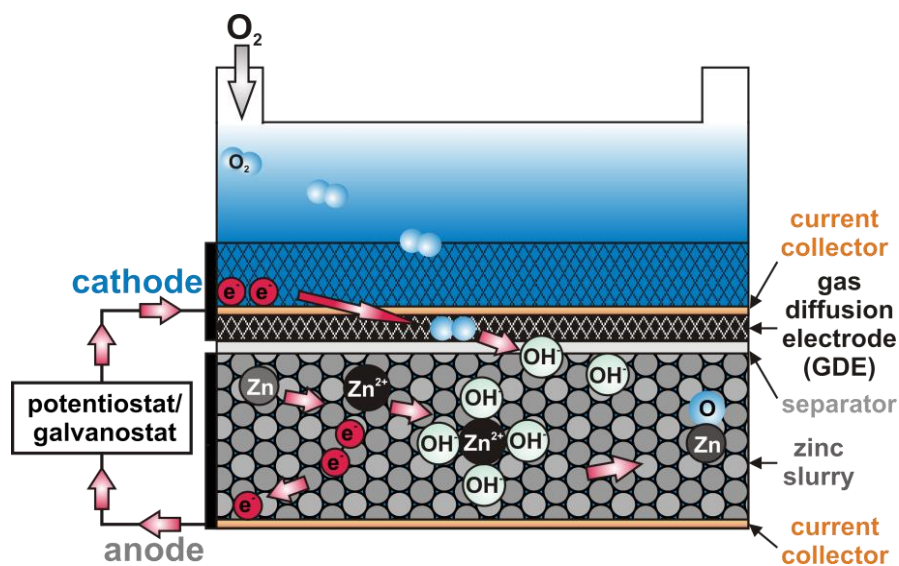

**Figure 1**: Schematic illustration of zinc air battery which is used for experiments.

The processes that occur during the cell discharge can be described with the following electrochemical reactions:

**Anode: Zinc Oxidation**

$$Zn + 4OH^- \rightarrow [Zn(OH)_4]^{2-} + 2e^- \qquad E^0 = -1.25 \text{ V vs. NHE}$$

**Elektrolyte:**

$$[Zn(OH)_4]^{2-} \rightarrow ZnO + H_2O + 2OH^-$$

**Cathode: Oxygen Reduction Reaction (ORR)**

$$O_2 + 2H_2O + 4e^- \rightarrow 4OH^- \qquad E^0 = +0.4 \text{ V vs. NHE}$$

**Overall reaction:**

$$2Zn + O_2 \rightarrow 2ZnO \qquad E^0 = +1.65 \text{ V}$$

Electrochemical reactions of a zinc air cell.[18]

However it has been observed that zinc particle with potassium hydroxide and commercial zinc slurry has a zinc coating activity to various metals. During the measurement with zinc air battery cell, a grey color coating was recognized on the copper current collector. After removing the copper from direct contact with the zinc particles the zinc coating disappears, this proves the reversibility of the zinc deposition. For more analysis the coating activity of zinc slurry is investigated with different materials: copper, silver and gold. The coating is analyzed with different methods: Scanning Electron microscope (SEM), Focused Ion Beam (FIB) and X-ray Photoelectron Spectroscopy (XPS). These investigations should clear if the coating on the surface is a pure zinc layer or not. A possible reaction will also be given for this activity.

## 3. Experimental

### 3.1. Materials

The commercial available zinc slurry from Grillo-Werke AG has a particle size of 25-60 μm and a purity of 99.9 %. For the self-prepared zinc slurry zinc powder is purchased from Alfa Aesar with 5-10 mesh and a purity of 99.8 % metals bases and the potassium hydroxide from Grüssing has a purity of 99.5 %. Sample platelets of copper, silver and gold are from Goodfellow with a purity of 99.9 %. Current collector with a thickness of approximately 1 mm and an area of around 1 cm$^2$.

### 3.2. Sample Preparation

Respective two specimens of the different materials are immersed in 2 ml of both, a commercial available and a self-made zinc slurry. The commercial available zinc slurry is used as received. For the preparation of the own zinc slurry, zinc powder is suspended in a weight ratio of 1 : 1 in a solution of 30 wt% potassium hydroxide in deionized water. After the specimens are immersed for 48 h in the respective slurry, they are carefully removed again and thoroughly cleaned with water, ethanol and acetone. Finally they are dried at room temperature for the following analyses.

### 3.3. Characterization

#### 3.3.1. Scanning Electron Microscope (SEM)

Scanning electron microscopy and associated energy dispersive x-ray spectroscopy (EDX) are simultaneously performed with a JSM 7000F model from Jeol to gain an element mapping from the surface to the inside of the samples. Therefore, the sample specimens are grinded plane on one side for the investigation of the thereby obtained cross-section.

#### 3.3.2. Focused Ion Beam (FIB)

For the analysis the SEM/FIB dual system of a FEI HELOIS NanoLab 600 is used. Therefore a cut is made into the sample with a gallium ion beam and an ion channeling contrast imaging as well as a SEM element mapping of the cross section is carried out. For the cutting, the sample is tilted at 52 ° and the surface at the destined incision edge is coated with platinum. The platinization serves to protect against contamination of the origin sample surface and ensures a sharp, straight cutting edge. The platinum deposition takes place in two steps, first by electron-plating, followed by ion plating. The electron plating protects the sample surface, but it only leads to an unstable platinum layer, the ion plating then leads to a stable platinum layer. Finally the cut is made by ablation with the gallium ion beam.

#### 3.3.3. X-ray Photoelectron Spectroscopy (XPS)

The XPS experiments were performed with an ESCALab MII by by Vacuum Generators, using Al-K$_\alpha$ radiation ($\hbar\omega$ = 1486.6 eV). The photoelectrons were detected by a 150°-type hemispherical analyzer at a pass energy of 20 eV in normal emission mode. For a quantitative analysis of the elemental composition the intensities of the Zn2p$_{3/2}$, Cu2p$_{3/2}$ and O1s peak were scaled with the photoemission cross sections by Yeh and Lindau [23]. For depth profiling experiments an Ar ion beam of 4.8 keV energy was used. The calibration of the ablation rate was performed according to the procedure described in [24].

## 4. Results and Discussion

In both cases, using the commercial zinc slurry as well as using the zinc slurry which is made of zinc powder and KOH solution, a metallic-silver, matt-shiny deposition forms as a thin layer on the test specimens, which consist of more noble metals than zinc, namely copper, silver and gold (Fig. 2). In all other cases, where less noble metals than zinc or electrically nonconductive materials are used, no deposition is formed.

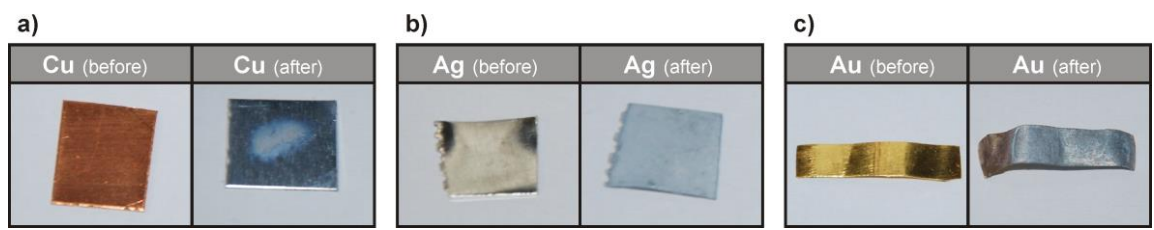

**Figure 2**: Sample platelets consisting of a) copper, b) silver and c) gold before and after immersion in zinc slurry or 48 h.

### 4.1. Scanning Electron Microscope (SEM)

In Fig. 3 the SEM images of the respective cross-section of the copper, silver and gold platelet (respectively grey panel) with the corresponding element mapping images (respectively colored panels) are shown. In the foreground of the grey panels, the plane grinded cross-section of the respective platelet can be recognized as the darkest surface. In the green panels the area where the spatially resolved K-line of respective copper, silver or gold are detected, is colored. In the blue panels a zinc layer on the surface is clearly recognizable by the coloring of the zinc K line at each sample. In the red panels the K line of oxygen is additionally detected in the areas in which the zinc plating is identified. Although it obviously seems to be a subsequent surface oxidation of the zinc deposition, this cannot definitely be clarified by means of this analytical method.

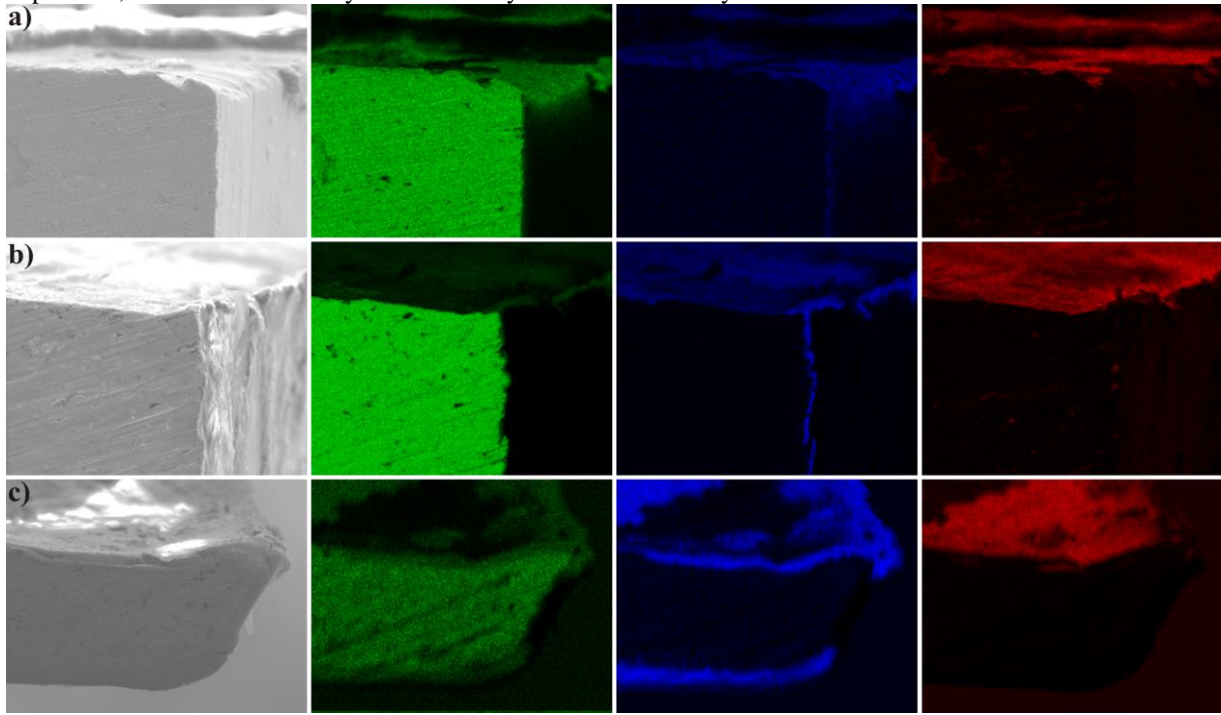

**Figure 3**: SEM images of the plane grinded cross section of the a) copper, b) silver and c) gold platelet with the green dyed K line signals of the respective metal, the blue dyed K line signals of zinc and the red dyed K line signals of oxygen.

### 4.2. Focused Ion Beam (FIB)

For a more detailed examination of the composition and the crystal structure, FIB is performed on the copper platelet sample. The necessary platinization, initially by electron plating and finally by ion plating, for the prevention of contaminations and alterations at the ablation area is represented by the SEM image in Fig. 4a. For the incision a gallium ion beam is used that burns directly into the protective platinum layer. In this way a sharp and straight cutting edge is obtained without any contamination or damage of the origin sample platelet surface, as can be seen in Fig. 4b. In Fig. 4c ion channeling is applied to obtain enhanced image contrast. Thereby, depending upon the different orientations of the crystal grains in the sample platelet, the gallium ions penetrate at different depth in the sample and thus produce an image contrast, which uncovers the crystallites with their different crystal orientation. Besides large crystallites in the inside, very small crystallites at the surface of the copper platelet can be seen. Also a triangular indent (marked by red arrow) can clearly be identified which is filled with these small crystallites. This is a strong hint for a supplementary deposition on the surface of the copper platelet with its well defined large crystallites and the random indent which already existed before.

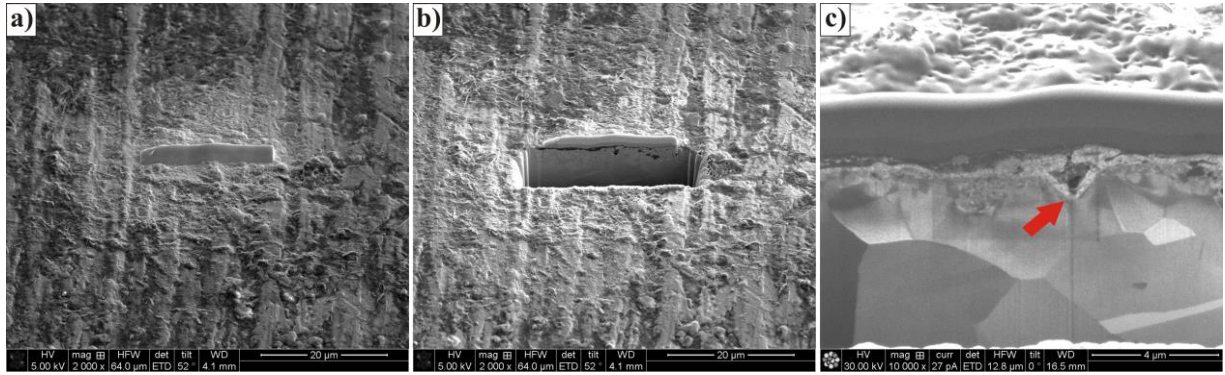

**Figure 4**: SEM images of the copper platelet a) after platinum deposition through electron plating and ion plating, b) after ablation by gallium ion beam and c) with Ion channeling contrast imaging of the cross section which uncovers a triangular indent in the copper platelet filled with small crystallites (marked by red arrow).

For the determination of the composition of the uncovered, fine grained deposition, SEM images with additional element mapping of the respective area are made. In Fig. 5a the indent on the surface of the copper platelet can clearly be identified again. In Fig. 5b the yellow dyed signals of the copper K lines properly describe the base body of the copper platelet with the indent. In Fig. 5c the purple zinc K lines specify the deposition which exactly fills out the indent. However, in d also green marked oxygen K lines can be seen in the same areas from where the zinc K lines originate. Indeed, it seems that the oxygen signals only belong to an oxidized state of the surface of the deposited zinc layer, but based on the results of this examination, no clear statement on the chemical state of the layer can be made.

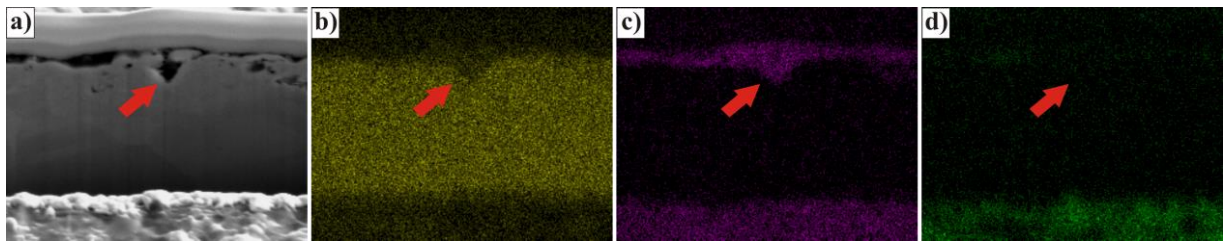

**Figure 5**: SEM images of a) the cross section with the triangular indent in the copper platelet (marked by red arrow) with b) yellow dyed copper K line signals, c) purple dyed zinc K line signals and d) green dyed oxygen K line signals.

## 4.3. X-ray Photoelectron Spectroscopy (XPS)

In order to clarify whether the deposited layer is a pure zinc layer, an oxidized zinc species or a mixture of both, XPS is performed. In Fig. 6 the relative amount (in at-%) of oxygen, zinc and copper is shown for different depths of the sample platelet, as obtained after stepwise ablation with an argon ion beam. At first, oxygen clearly predominates over the amount of zinc atoms until a depth of around 1 nm. This excess could be ascribed to adsorbed oxygen and to hydroxyl groups besides oxy groups on the surface of the zinc deposition. Then, this ratio is inverted by a fast decrease of oxygen whereas zinc further increases until a maximum at 5 nm from which it only decreases slowly until zinc and oxygen reach their minimum at about 40 nm where, in turn, copper reaches its maximum. So oxygen exists in the whole region of the deposition. This may be explained by an at least partial oxidation of the surface of the copper platelet before the deposition of the zinc layer. However, according to the line shape of the $Cu2p_{2/3}$ peak in XPS (not shown) the presence of CuO can be excluded. Therefore the exact bonding states of zinc have to be analyzed.

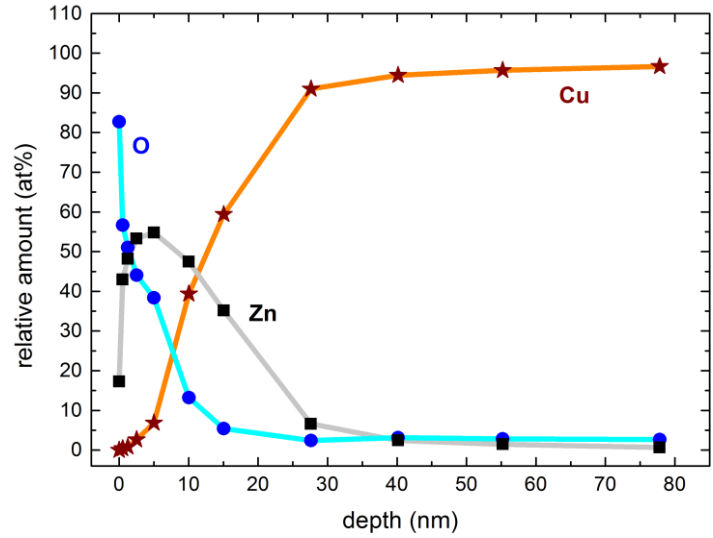

**Figure 6**: Elemental depth profiles of the sample platelet for the relative amount (in at-%) of O, Zn and Cu as determined by a combination of XPS measurements and the surface ablation of the sample via Ar ion etching.

To reveal the exact bonding states of zinc, the Auger Zn $L_3M_{4,5}M_{4,5}$ lines are analyzed. According to Deroubaix et al. [25] the kinetic energies of these Auger electron differ by approx. 4 eV for $Zn^0$ and $Zn^{2+}$. Fig. 7 shows the XPS-derived Auger Zn $L_3M_{4,5}M_{4,5}$ spectra in dependence on the overall ablation of the initial surface. At the beginning of the ablation only the Auger lines of the second oxidation state are present, and thus the surface consists of ZnO. Down to 5 nm a change in the oxidation step is identified, wherein the portion of the metallic state is increased, and finally completely passes on elemental zinc at 10 till 15 nm. Most probably the outer zinc oxide layer is formed by oxidation in air after the deposition. By the fact of the underlying zinc layer, it unequivocally can be concluded that elemental zinc is deposited on the surface of the copper sample platelet.

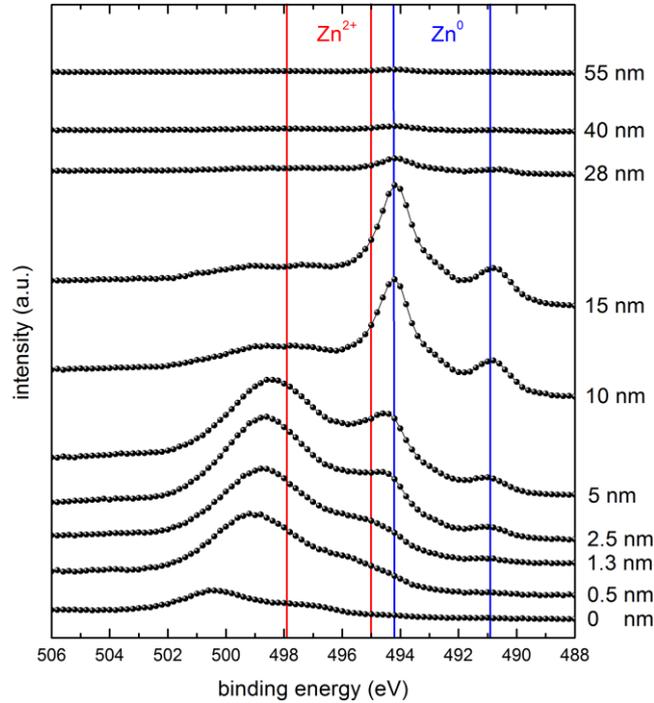

**Figure 7**: Depth profile of the sample platelet with the oxidation states of zinc determined by a combination of the intensity of the Auger Zn $L_3M_{4,5}M_{4,5}$ lines of $Zn^0$ and $Zn^{2+}$ at different depths after the ablation with an argon ion beam. The red and blue lines refer to the positions of $Zn^{2+}$ and $Zn^0$, respectively, with the kinetic energies according to Ref. [25] transformed to binding energies (as usually displayed in XPS data).

## 5. Reaction mechanism

In the following an additional experiment is described in order to give a possible explanation for the observations and the occurrence of the zinc deposition. Based on the aforementioned experiments a zinc slurry is prepared utilizing an excess of the aqueous potassium hydroxide solution (30 wt%) resulting in a sediment of zinc powder and a supernatant of potassium hydroxide solution (Fig. 8a). Under these conditions a small amount of zinc is slowly oxidized accompanied by hydrogen evolution and the formation of the tetrahydroxozincate complex.[18] Then a copper platelet is immersed in the solution (Fig. 8b). At these conditions no zinc coating is observed, even after several days. In a next step the copper platelet is additionally immersed in the zinc powder sediment (Fig. 8c). Since copper is a much more noble metal compared to zinc, it subtracts electrons from zinc when in contact (Fig. 8c). After 24 h a thin layer of zinc forms on the whole part of the copper platelet which is immersed into the solution (Fig. 8d). In the following possible reactions are discussed, quoted standard potentials are taken from ref. [26].

First suggestion for this reaction is given in following mechanism:

$$Cu_2O + 2OH^- + H_2O \rightarrow 2Cu(OH)_2 + 2e^- \qquad E^0 = +0.80 \text{ V vs. NHE}$$
$$[Zn(OH)_4]^{2-} + 2e^- \rightarrow Zn + 4OH^- \qquad E^0 = -1.25 \text{ V vs. NHE}$$

On the surface of copper there will be oxidation of copper. And the tetrahydroxozincate complex is chemical reduced with the electron of the copper oxidation. If the reaction proceeds through the oxidation of the metal then it should also happen in case of gold and silver, which is unlikely for gold. Another reason is that the reaction should already start without the direct contact between zinc and copper. However, the zinc deposition occurs only with the direct contact of both metals.

The second suggestion is through the hydrogen oxidation which is mentioned below:

$$H_2 + 2OH^- \rightarrow 2H_2O + 2e^- \qquad E^0 = +0.83 \text{ V vs. NHE}$$
$$Zn^{2+} + 2e^- \rightarrow Zn \qquad E^0 = -0.76 \text{ V vs. NHE}$$

Due to the electron loading on copper, hydrogen is formed. This hydrogen can lead the reduction of zinc ion to zinc. However, the standard potentials reveal that this reaction can't occur spontaneously.

The third suggestion is the most plausible reaction. The electron, which is pulled from zinc to copper, is loading up the copper negative. If the tetrahydroxozincate complex, which also negative loaded, converges to copper the solvation shell is disintegrated and it forms to zinc ions and hydroxide ions. Apparently, the dissolved zinc ions can be reduced to metallic zinc by accepting excess electrons from copper which primarily originate from the zinc powder bulk. The possible reaction can be described below:

$$[Zn(OH)_4]^{2-} \rightarrow Zn^{2+} + 4OH^-$$
$$Zn^{2+} + 2e^- \rightarrow Zn \qquad E^0 = -0.76 \text{ V vs. NHE}$$

However, in this way, only a very thin layer can be obtained which doesn't grow any further, even after several weeks. Obviously, a dynamic equilibrium is reached at a certain thickness of the deposited zinc layer where the electron-donating force of the zinc layer is equal to the one of the zinc powder bulk. Finally the copper platelet is disconnected from the zinc powder sediment (Fig. 8e) and after 24 h the complete zinc layer, which is immersed in the solution is decomposed again (Fig. 8f), due to the slow oxidation of zinc together with the evolution of hydrogen and the lack of excess electrons coming from the disconnected zinc powder bulk. Overall, it is a movement of elemental zinc from the powder bulk over the tetrahydroxozincate complex to the surface of the copper platelet.

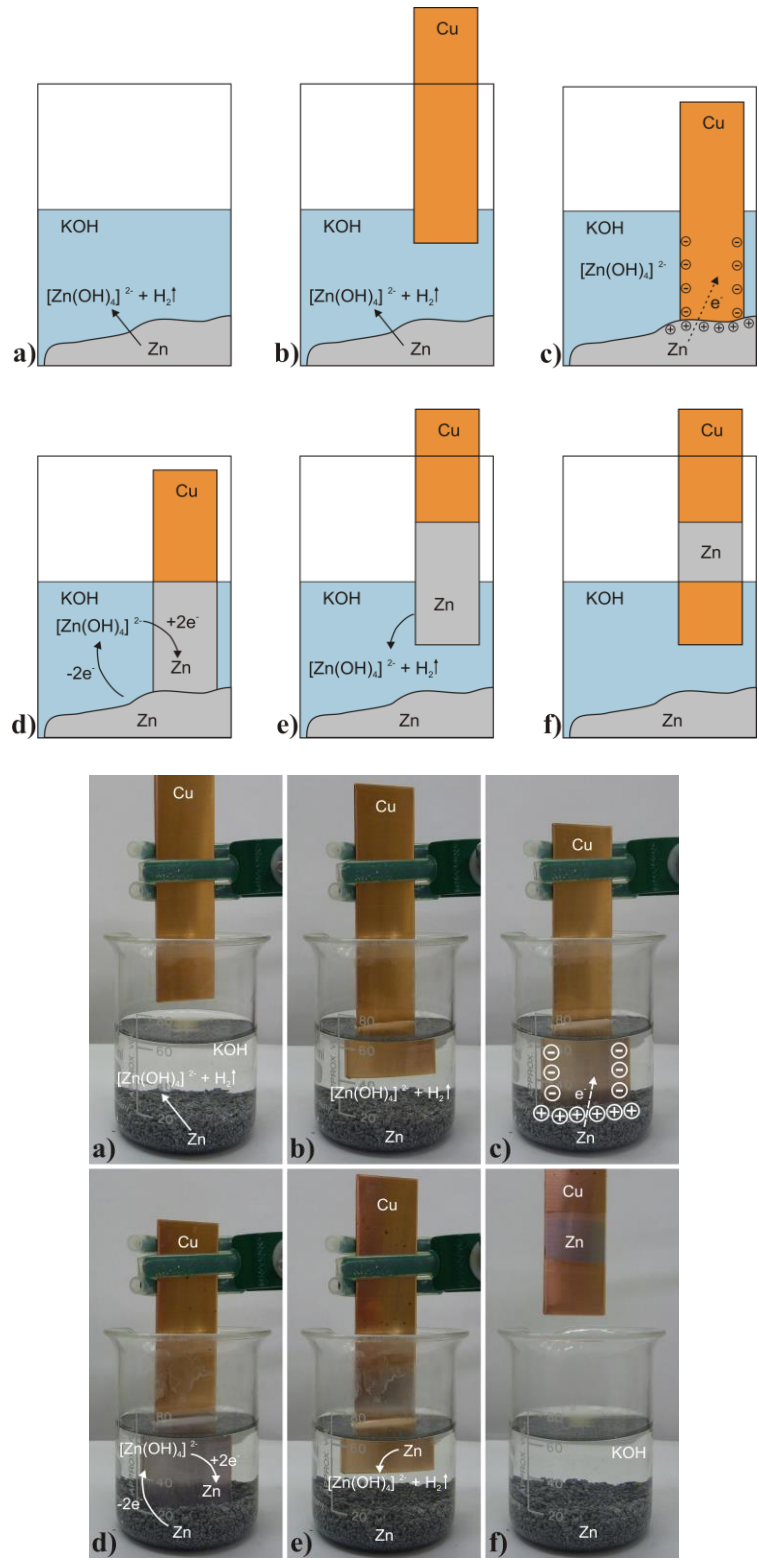

**Figure 8**: Schematic illustration of a) the slow electrolysis of the zinc powder sediment b) putting copper platelet in the electrolyte during the slow electrolysis of the zinc powder sediment, c) the subtraction of electrons from zinc to copper, d) the zinc layer deposition on the copper platelet over tetrahydroxozincate complex as a transition state, e) disconnection of the zinc plated copper platelet with electrolysis of the zinc layer and f) dissolution of the zinc layer, which stays immersed in the solution due to slow electrolysis of the zinc.

## 6. Conclusion

The zinc air battery shows an unwanted electro less zinc plating on the current collector during the measurement. For the analysis, a commercial zinc slurry is used to investigate the zinc plating on copper, silver and gold. All three materials show clearly silver colored plating on their surfaces. For the surface examination, all three samples are studied by means of SEM including element mapping. The investigation confirms the zinc plating on the surface, but due to surface oxidation of the deposited zinc it is still not clear if metallic zinc or zinc oxide is deposited. For further research, the copper sample is examined with FIB and XPS method. With FIB the surface of the zinc plated copper sample is cut and the edge is protected with platinum. With the so-obtained sharp and straight cutting edge, the composition of the surface is analyzed with SEM. The result does not clearly differentiate whether it is a zinc layer or a zinc and oxide layer on the surface of the copper. For this purpose XPS is used to reveal the chemical state of the layer. When combined with Argon ion etching the surface is removed stepwise and the surface is analyzed in terms of its chemical composition and the chemical states of the detected elements. The XPS measurement shows first the Auger lines of the second oxidation state of Zn. But with further ablation the Auger lines of the metallic Zn state increase. This research of the silver colored plating on the material surface proves that a zinc deposition takes place on the surface. The air contact of the samples with a small layer of zinc oxide is located on the surface through the oxidation process.

Finally a possible reaction is given for this process. In an alkaline solution the zinc is oxidized and it gets slowly in solution as tetrahydroxozincate complex. While putting the copper in contact with the zinc powder, the electrons are transferred from zinc to copper. After several time the dissolved zinc ions are reduced to zinc by accepting excess electrons from copper and form a zinc deposition on the copper surface. In the case that the copper is disconnected from the zinc, the initially formed zinc coating, which stays immersed in the solution, disperses. The zinc deposition in alkaline solution on copper is overall a movement of elemental zinc from the powder bulk over the tetrahydroxozincate complex to the surface of the copper platelet.

## 7. Acknowledgement

Co-funding this work through the project ´´ZnPlus – Rechargeable zinc-air batteries for energy storage`` (support code: 03ESP217F) in the framework of BMWi-joint (Federal Ministry for Economic Affairs and Energy) research project is gratefully acknowledged.
(http://forschung-energiespeicher.info/en/projektschau/gesamtliste/projekt-einzelansicht/95/Zink_Luft_Akkus_fuers_Netz/    last visited: 17.01.2017)